**Optically-driven nuclear-spin-selective Rabi oscillations of single electron spins in diamond**

D. Andrew Golter and Hailin Wang

Department of Physics and Oregon Center for Optics

University of Oregon, Eugene, Oregon 97403, USA

Abstract

Optically-driven Rabi oscillations of single electron spins in a diamond nitrogen vacancy center are realized with two Raman-resonant optical pulses that are detuned from the respective dipole optical transitions. The Rabi oscillations are nuclear-spin selective and are robust against rapid decoherence of the underlying optical transitions, opening the door to generating phonon-mediated spin entanglement in an optically-driven spin-phonon system. A direct comparison between the Rabi oscillations and simulated Raman adiabatic passage along with a detailed theoretical analysis provide a wealth of information on the coherent spin dynamics and in particular reveal that decoherence and environmental fluctuations associated with the optical transitions affect the fidelity of the Rabi oscillations through relatively large jumps in the transition frequencies.

Negatively-charged nitrogen vacancy (NV) centers in diamond provide a paradigmatic system for pursuing quantum control of electron and nuclear spins. NV centers feature extraordinarily long decoherence times for electron and nuclear spins, along with high fidelity optical state preparation and readout[1-5]. A well-established approach for coherent spin manipulation in NV centers is to drive the relevant electron spin transitions with microwave (MW) radiation[6,7]. The exquisite control of electron and nuclear spin dynamics in NV centers has also led to the realization of quantum state transfer between electron and proximal nuclear spins[8], enabling the use of nuclear spins as quantum memories.

Recent advances in quantum optomechanics have also stimulated strong interest in coupling electron spins to high-Q mechanical vibrations to realize spin entanglement as well as spin-photon quantum interfaces[9-12]. A highly promising approach proposed recently is to exploit strong excited-state electron-phonon interactions in NV centers to induce coherent spin dynamics through phonon-assisted optical transitions[12]. High-Q nanomechanical oscillators of single crystal diamond have been fabricated recently[13]. An optically-driven spin-phonon system resembles the well-established atomic system of trapped ions, providing an excellent platform for exploring foundational quantum phenomena as well as scalable quantum information processing in a solid state environment. For the pursuit of the optically-driven spin-phonon system and more generally for optical control of electron spins, a fundamental issue is how to control coherent spin dynamics through optical transitions, while avoiding the dipole decoherence of the optical transitions. To take advantage of the unique properties of nuclear spins, it is also important that the optically-driven spin dynamics be nuclear-spin selective.

In this paper, we report experimental demonstration of optically-driven Rabi oscillations of a single electron spin in a NV center by using two Raman-resonant optical pulses that are detuned from the respective dipole optical transitions. We show that the Rabi oscillations, with a decay time over 1 µs, are robust against rapid decoherence of the underlying optical transitions and also depend on the orientation of the adjacent $^{14}$N nuclear spin. In addition, we have made a direct comparison between the optically-driven Rabi oscillations and stimulated Raman adiabatic passage (STIRAP) in the settings of a single experiment. The experimental results agree well with a theoretical analysis based on the optical Bloch equations, providing significant insights into the optically-driven spin dynamics, especially on how decoherence processes associated with the optical transitions affect the fidelity of the Rabi oscillations.

The electronic ground state of the negatively charged NV center is a spin triplet, characterized by $m_s=0$ and $m_s=\pm 1$ states[14]. Spin conserving optical transitions can take place between the ground states and the six excited states[15, 16]. Figure 1a shows a $\Lambda$-type three-level system, with spin states $m_s=\pm 1$ as the two lower levels and with excited state $A_2$ as the upper level. The $A_2$ state couples to the $m_s=+1$ and $m_s=-1$ states via σ- and σ+ polarized light, respectively. A static magnetic field can be used to induce a Zeeman splitting between the $m_s=\pm 1$ states. The three energy levels form a nearly closed system, with a small probability for an electron in state $A_2$ to decay nonradiatively into the $m_s=0$ state[17]. The electron spin in the NV center also couples to the $^{14}$N nuclear spin, with a hypefine splitting of 2.2 MHz[14]. For the $m_s=\pm 1$ states, each electron spin state splits into three hyperfine states, corresponding to nuclear spin projection with $m_n=-1, 0, +1$ (see Fig. 1a).

In the limit that two simultaneous Raman-resonant incident optical fields coupling to the two dipole transitions in the $\Lambda$-system are sufficiently detuned from the respective dipole transitions, the excited state dynamics can adiabatically follow the incident optical fields as well as the dynamics of the two spin states. In this limit, the $\Lambda$-type system in Fig. 1a can in principle be reduced to a two-level spin system, with the effective Rabi frequency for the optically-driven spin transition given by[18]:

$$\Omega_R = \Omega_+\Omega_-/(2\Delta) \tag{1}$$

where $\Omega_+$ and $\Omega_-$ are the Rabi frequencies for the respective dipole optical transitions and $\Delta$ is the average detuning of the optical fields from the dipole transitions. In this case, two simultaneous Raman-resonant optical fields can drive a Rabi oscillation of the electron spin. Optically-driven spin dynamics can also take place via a coherent population trapping (CPT) or dark state[19]. In the limit that the $\Lambda$-type three-level system is optically pumped or driven into the CPT state,

$$|\psi_d> = [\Omega_-(t)|+> - \Omega_+(t)|->]/\sqrt{\Omega_+^2(t)+\Omega_-^2(t)}, \tag{2}$$

the evolution of the spin system can be controlled adiabatically by varying the relative amplitude of the incident optical fields through the STIRAP process. CPT, including nuclear-spin-selective CPT, has been demonstrated in NV centers[20-23]. All optical control of electron spins via a dark state in NV centers has been explored recently by tuning a NV center to an excited-state spin anti-crossing[24].

Our experimental studies were carried out in a type IIa high purity diamond at low temperature (T ≈ 5 K) in a close-cycled optical cryostat. A solid immersion lens was etched into the surface of the diamond with a focused ion beam in order to improve the collection efficiency. A small magnetic field was applied to induce a Zeeman slitting, $\omega_B$ = 150 MHz. A green 532 nm diode laser provided off-resonant excitation of the NV center. A 637 nm tunable ring laser was used for the two nearly resonant incident optical fields. Optical pulses of and relative detuning between these two fields were generated with acousto-optic modulators. An additional diode laser, also at 637 nm, was used for spin readout through resonant optical excitation. Detailed information of the NV center used has been presented in an earlier study on CPT of electron spins[23]. Additional photoluminescence excitation (PLE) experiments show that the $A_2$ transitions are circularly polarized with an extinction ratio near 10:1.

As illustrated in the pulse sequence for the experiment on optically-driven Rabi oscillations (Fig. 1b), the single NV center was first initialized with a green laser pulse into the $m_s$=0 state. A MW π-pulse then prepared the NV center in the $m_s$=-1 state, with random nuclear spin orientation unless otherwise specified. In the manipulation step, two simultaneous square-shaped optical pulses that are Raman resonant with the $m_s$=±1 states and detuned from the $A_2$ transition drove the coherent evolution of the electron spin. The two optical fields have nearly equal peak intensity and are oppositely circular polarized. For the spin detection, the spin population in either the $m_s$=+1 or $m_s$=-1 state was measured with a two-step PLE-based process. First, a MW π-pulse drove the relevant spin population into the $m_s$ = 0 state. Second, the transition between the $m_s$ = 0 state and the $E_y$ excited state was resonantly excited, with the resulting fluorescence measured at the lower energy phonon sideband.

Figure 1c shows that the electron spin population oscillates with the duration of the incident optical pulses, where Δ is set to 1.5 GHz and the optical pulses are Raman resonant with the spin states associated with $m_n$=0 (i.e., $\delta_0$=0 with $\delta_n$ being the detuning from the respective Raman resonance for given $m_n$). As expected from Eq. 1, the period of the Rabi oscillation determined from the experiments is proportional to Δ; the effective Rabi frequency, $\Omega_R$ is proportional to $\sqrt{I_+ I_-}$ (see Fig. 1d), where $I_+$ and $I_-$ are the peak intensities for the respective optical pulses. These results confirm that driven by the detuned and Raman-resonant optical fields, the Λ-type system of the NV center behaves like an effective two-level spin system. As shown in Fig. 1c, the Rabi oscillations are robust against the rapid dipole decoherence associated with the underlying optical transitions, with a decay time of 1.3 µs, compared with an excited state lifetime of 11.5 ns (corresponding to an intrinsic linewidth of 13 MHz) and an absorption linewidth of 500 MHz for single NV ceners[25-27]. Decay of the Rabi oscillations is in large part due to nuclear spin diffusion, which can be overcome with dynamic decoupling[28]. Decay mechanisms related to the optical transitions will be discussed in detail later.

To demonstrate the dependence of the optically-driven Rabi oscillations on the orientation of the $^{14}$N nuclear spin associated with the NV center, we measured the electron spin evolution as a function of the optical pulse duration at various frequency differences between the two incident optical fields. For clarity, we have included in Fig. 2a the spectral response obtained in an earlier CPT experiment performed on the same NV center, where the three dips correspond to the Raman resonant condition associated with $m_n$=-1, 0, +1, respectively[23]. As shown in Fig. 2, Rabi oscillations of the electron spin are observed when the frequency difference is at the dips of the CPT spectral response, i.e. when the $m_n$-dependent Raman resonant condition is satisfied. The oscillation, however, vanishes when the frequency difference is midway between the two nearby dips (see Fig. 2b). As long as $\Omega_R$ is small compared with the relevant hyperfine splitting, the optically-driven Rabi oscillation is nuclear-spin selective, which is essential for the use of nuclear spins associated with the NV center in processes such as quantum memories.

For the experiments shown in Figs. 1 and 2, the nuclear spin orientation is random and the Rabi oscillations are nuclear spin selective. Thus, only 1/3 of the total electron population is involved in the Rabi oscillation. For Rabi oscillations with perfect fidelity, the minimum normalized fluorescence for the $m_s$ = -1 state is 2/3 and the maximum normalized fluorescence for the $m_s$=+1 state is 1/3. With the relatively long Rabi period, the fidelity of the Rabi oscillation shown in the top trace of Fig. 1c is estimated to be 0.9. Note that the MW π-pulse used for the $m_s$=+1 state in the spin detection step was not ideal, leading to an overall decrease in the measured electron population (the bottom trace in Fig. 1c). The decay of the electron population in the $m_s$=-1 state shown in Fig. 2b provides a separate measure of the excitation of the $A_2$ state and the subsequent optical pumping, i.e., decay of the electron from the $A_2$ state to the $m_s$=+1 as well as the $m_s$=0 state. The slow decay shown in Fig. 2b indicates that the optical pumping has a relatively minor contribution to the decay of the Rabi oscillation.

We have used the optical spin rotation to determine the dephasing time of the electron spin. For these experiments, the electron population was initialized into the $m_s$=-1 state with a given nuclear spin orientation, $m_n$, using a nuclear-spin-selective MW π-pulse. Two Raman-resonant optical pulse pairs then induced π/2 rotations separated by time τ (see Fig. 3a). After the second π/2 rotation, the population in the $m_s$=+1 state was measured as before. Figures 3b-d show the electron spin precessions obtained with $m_n$=0, -1, +1 respectively, with $\delta_0$=1.4 MHz. The solid lines in these figures show a numerical fit to $\exp[-(\tau/T_2^*)^2]\cos(2\pi\delta_n)$ with a dephasing time $T_2^*$ = 1 μs, in agreement with separate Ramsey fringe experiments performed with MW-driven π/2 pulses. In Fig. 3e, the electron population was initialized into the $m_s$=-1 state with a random nuclear spin orientation. The solid line in Fig. 3e shows the sum of the three numerical fits obtained for individual $m_n$, with no adjustable parameter except for an overall scale factor. These experiments show that the optically-driven spin rotation is as effective as the conventional MW-driven spin rotation.

Optically-driven coherent spin evolution can also take place via STIRAP. For a detailed understanding of the coherent spin dynamics, we have devised a method to perform and compare both Rabi oscillations and STIRAP in a single experiment. The pulse sequence used is similar to that in Fig. 1, but now we tailor the temporal lineshape of the Raman-resonant optical pulses and delay the two pulses relative to each other. As shown in Fig. 4a, the rising edge of the $\Omega_-$ pulse and the trailing edge of the $\Omega_+$ pulse are characterized by time $t_{\text{rise}}$. The separation between the rising edge of the $\Omega_+$ pulse and the trailing edge of the $\Omega_-$ pulse is defined as $T$. For Figs. 4b-e (left column), the population in the $m_s$ =+1 state, following the application of the two optical pulses, was measured as a function of $T$, with Δ=0.9 GHz and with other conditions remaining the same as those in Fig. 1.

Figure 4 shows the close connection and the transition between the optically-driven Rabi oscillation and the STIRAP. With $t_{\text{rise}}$=1.2 μs, STIRAP occurs as the trailing edge of the $\Omega_+$ pulse overlaps with the rising edge of the $\Omega_-$ pulse (see Fig. 4b). The Rabi oscillations take place when the peak amplitudes of two optical pulses overlap in time. With decreasing $t_{\text{rise}}$ (see Figs. 4c-d), the adiabatic condition breaks down and the STIRAP transitions into Rabi oscillations. When the two optical pulses are nearly square-shaped (see Fig. 4e), the Rabi oscillations become nearly symmetric with respect to $T$=1.5 μs, at which the two pulses overlap completely. This direct comparison between Rabi oscillations and STIRAP indicates that adiabatic evolution of the dark state can only take place on a timescale slower than the Rabi period (as it occurs when the two optical pulses are at their peak amplitude). For fast optically-driven spin transition, Rabi oscillations are thus preferred over STIRAP.

For a theoretical description of the experiments, we have modeled the spin dynamics with optical Bloch equations for a Λ-type three-level system[29]. The right column of Figs. 4b-e plots the theoretical calculation under the conditions of the experiments in the left column, where we have used $T_2$= 200 μs for the electron spin and an intrinsic decoherence rate $\gamma/2\pi$ = 7 MHz for the $A_2$ transitions, with all other parameters, including $T_2^*$ = 1 μs, derived from the experiments. Fluctuations or jumps of the optical transition frequency due to re-initialization of the NV center by a green laser were modeled as an inhomogeneous broadening, with a linewidth of 500 MHz as determined from the PLE spectrum[25, 27]. Figure 4 shows an overall good agreement between the theory and experiment[30]. The calculation in Fig. 4 also indicates that the slight asymmetry in Fig. 4e provides a sensitive measure of optical pumping induced by the excitation of the $A_2$ state.

As discussed in detail in the supplement[30], the theoretical analysis describes well the decay of the Rabi oscillations in Fig. 1 and Fig. 4. Dipole optical transitions affect the fidelity of the coherent spin process through fluctuations in the optical transition frequencies[30]. Relatively large spectral jumps lead to corresponding changes in $\Omega_R$ and can also induce increased excitations of the $A_2$ state. With the dipole detuning used and without the spectral fluctuations, decoherence of the optical transitions has a negligible effect on the fidelity of the Rabi oscillations[30]. The spectral fluctuations, however, can be suppressed, if deshelving of NV centers is performed through resonant optical excitation of neutral NV centers, along with the use of resonant excitation for spin initialization[31].

In summary, we have demonstrated optically-driven nuclear-spin-selective Rabi oscillations of single electron spins in diamond. With adequate dipole detuning and with spectral fluctuations of the NV optical transitions suppressed, high fidelity Rabi oscillations can be achieved, with negligible effects from decoherence associated with the underlying optical transitions. These remarkable coherent spin phenomena open the door to using an optically-driven spin-phonon system for realizing a trapped-ion system in a solid and especially for generating phonon-mediated entanglement of electron spins in diamond.

This work is supported by NSF under award No. 1005499.

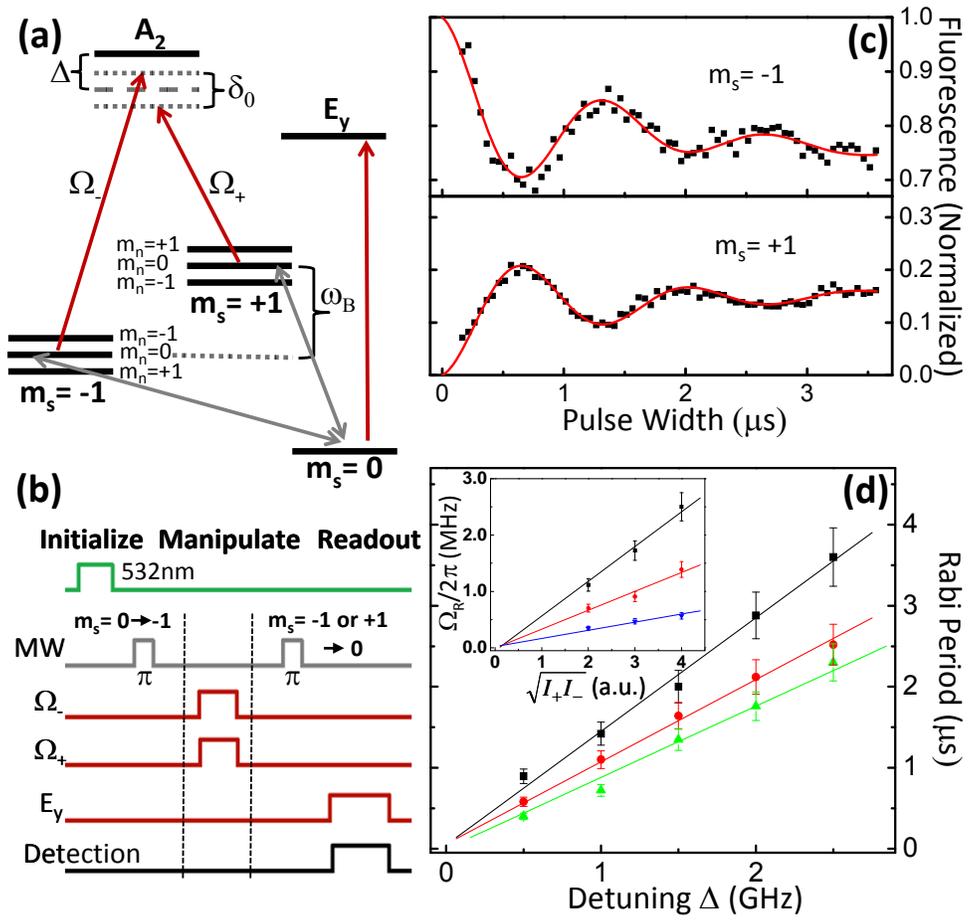

FIG. 1 (color online). (a) Energy level structure of a NV center, including the ground-state spin triplet, hyperfine splitting due to the $^{14}$N nuclear spin, and the relevant excited states. $\Omega_-$ and $\Omega_+$ are the optical Rabi frequencies. (b) Pulse sequence used for optically-driven Rabi oscillations. (c) Optically-driven Rabi oscillations of an electron spin. The fluorescence measures the population in the $m_s=-1$ and $m_s=+1$ states. The solid line is a numerical fit to a damped oscillation with an added slope due to optical pumping. (d) Period of the Rabi oscillations as a function of detuning, $\Delta$, for three different optical intensities. Inset: Effective Rabi frequency as a function of the intensity for three different $\Delta$.

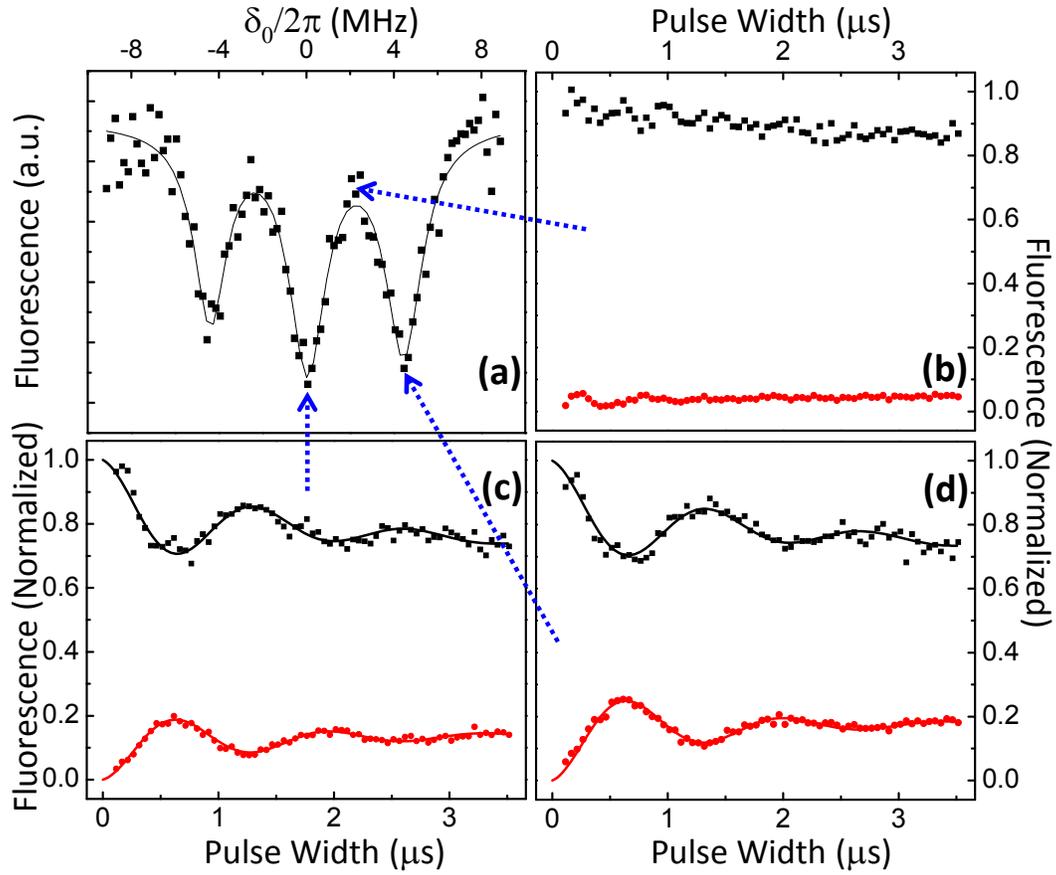

Fig. 2 (color online). Nuclear-spin-selective Rabi oscillations. For clarity, the detunings are indicated in (a), the spectral response of the nuclear-spin-dependent CPT. In (b)-(d), the populations in the $m_s=-1$ (top traces) and $m_s=+1$ (bottom traces) states are shown for three different detunings between the two incident optical pulses. Solid lines in (c) and (d) are numerical fits to damped oscillations.

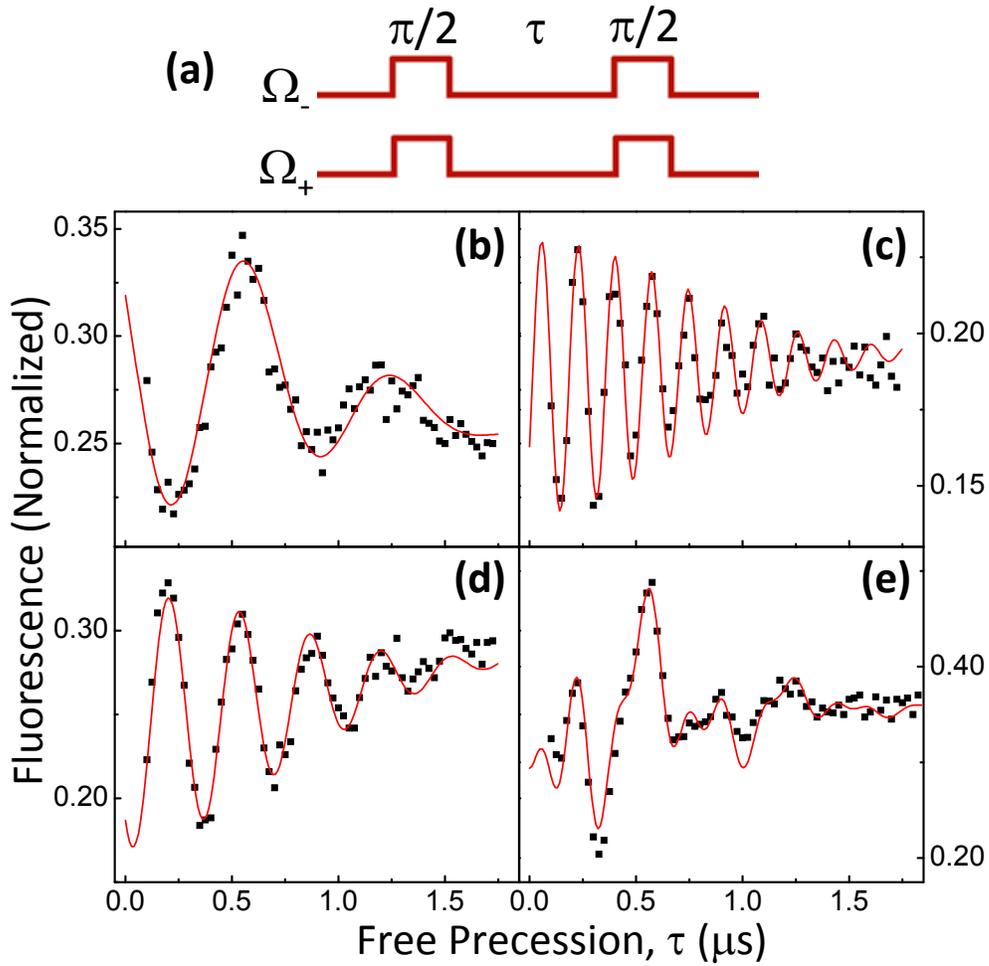

Fig. 3 (color online). (a) Pulse sequence used for the Ramsey fringe experiment. (b)-(e) Free induction decays of an electron spin with $\Omega_R/2\pi=2.5$ MHz and $\Delta=1$ GHz. For (b), (c), and (d), nuclear-spin-selective MW $\pi$-pulses were used to prepare the electron in the $m_s=-1$ and $m_n=0$, -1, and +1, hyperfine states, respectively. For (e), the electron was prepared in the $m_s=-1$ state with random nuclear spin orientation. Solid lines are numerical fits as discussed in the text.

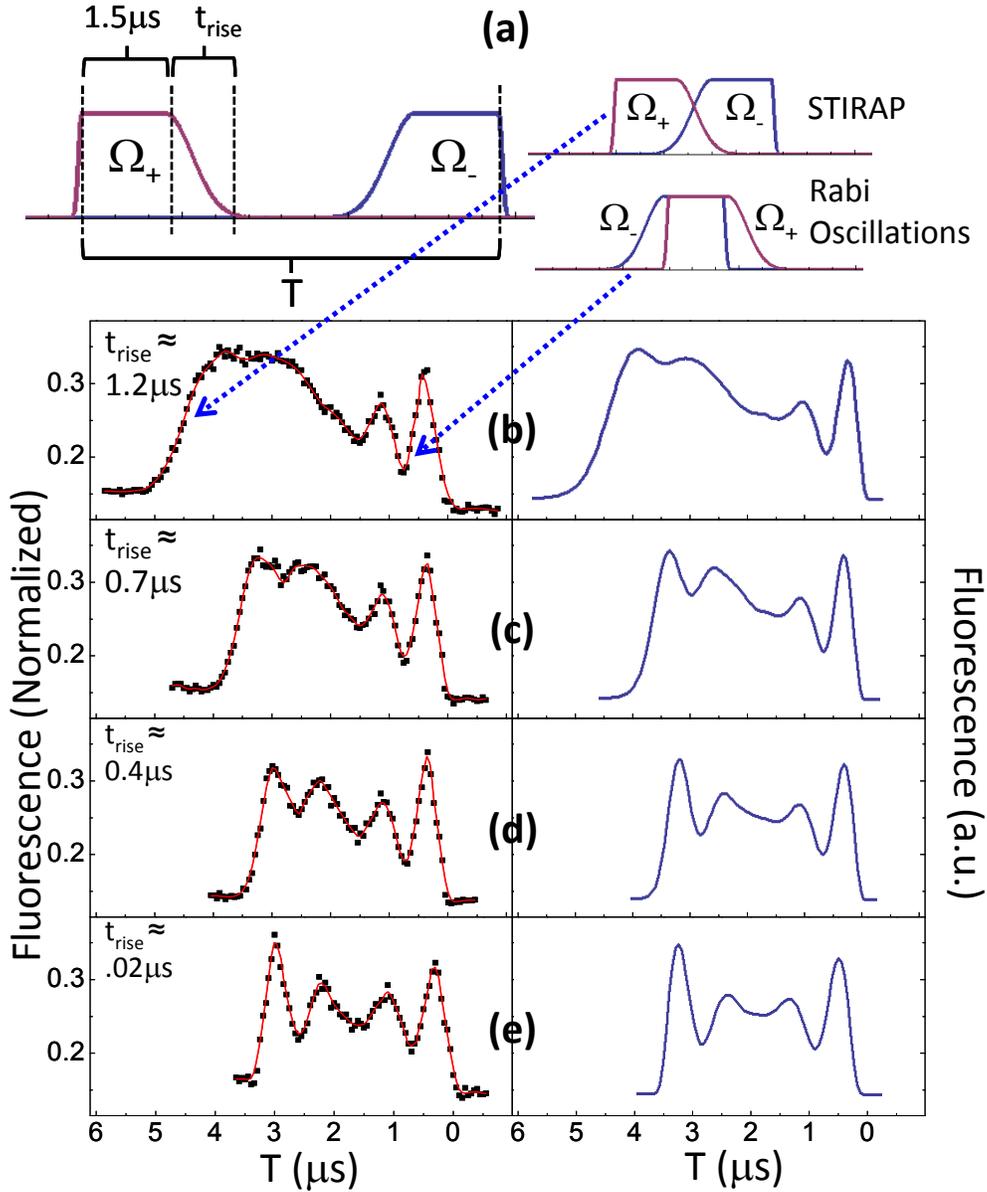

Fig. 4 (color online). (a) The temporal lineshapes of the two optical pulses used for the experiment, for which the population in state $m_s$=+1 was measured as a function of delay, $T$, between the two pulses. (b)-(e) Left column: experimental results obtained with different $t_{rise}$, as indicated in the figure. Solid lines are guide to the eye. Right column: theoretical calculations using parameters of the experiments in the left column.

**References:**

[1]  G. Balasubramanian, P. Neumann, D. Twitchen, M. Markham, R. Kolesov, N. Mizuochi, J. Isoya, J. Achard, J. Beck, J. Tissler, V. Jacques, P. R. Hemmer, F. Jelezko, and J. Wrachtrup, Nature Materials **8**, 383 (2009).
[2]  P. C. Maurer, G. Kucsko, C. Latta, L. Jiang, N. Y. Yao, S. D. Bennett, F. Pastawski, D. Hunger, N. Chisholm, M. Markham, D. J. Twitchen, J. I. Cirac, and M. D. Lukin, Science **336**, 1283 (2012).
[3]  L. Robledo, L. Childress, H. Bernien, B. Hensen, P. F. A. Alkemade, and R. Hanson, Nature **477**, 574 (2011).
[4]  L. Jiang, J. S. Hodges, J. R. Maze, P. Maurer, J. M. Taylor, D. G. Cory, P. R. Hemmer, R. L. Walsworth, A. Yacoby, A. S. Zibrov, and M. D. Lukin, Science **326**, 267 (2009).
[5]  P. Neumann, J. Beck, M. Steiner, F. Rempp, H. Fedder, P. R. Hemmer, J. Wrachtrup, and F. Jelezko, Science **329**, 542 (2010).
[6]  F. Jelezko, T. Gaebel, I. Popa, A. Gruber, and J. Wrachtrup, Physical Review Letters **92**, 076401 (2004).
[7]  G. D. Fuchs, V. V. Dobrovitski, D. M. Toyli, F. J. Heremans, and D. D. Awschalom, Science **326**, 1520 (2009).
[8]  M. V. G. Dutt, L. Childress, L. Jiang, E. Togan, J. Maze, F. Jelezko, A. S. Zibrov, P. R. Hemmer, and M. D. Lukin, Science **316**, 1312 (2007).
[9]  P. Rabl, P. Cappellaro, M. V. G. Dutt, L. Jiang, J. R. Maze, and M. D. Lukin, Physical Review B **79**, 041302 (2009).
[10]  K. Stannigel, P. Rabl, A. S. Sorensen, P. Zoller, and M. D. Lukin, Physical Review Letters **105**, 220501 (2010).
[11]  S. Kolkowitz, A. C. B. Jayich, Q. P. Unterreithmeier, S. D. Bennett, P. Rabl, J. G. E. Harris, and M. D. Lukin, Science **335**, 1603 (2012).
[12]  A. Albrecht, A. Retzker, F. Jelezko, and M. B. Plenio, New Journal of Physics **15**, 083014 (2013).
[13]  M. J. Burek, D. Ramos, P. Patel, I. W. Frank, and M. Lončar, Applied Physics Letters **103**, 131904 (2013).
[14]  M. W. Doherty, F. Dolde, H. Fedder, F. Jelezko, J. Wrachtrup, N. B. Manson, and L. C. L. Hollenberg, Physical Review B **85**, 205203 (2012).
[15]  A. Batalov, V. Jacques, F. Kaiser, P. Siyushev, P. Neumann, L. J. Rogers, R. L. McMurtrie, N. B. Manson, F. Jelezko, and J. Wrachtrup, Physical Review Letters **102**, 195506 (2009).
[16]  G. D. Fuchs, V. V. Dobrovitski, R. Hanson, A. Batra, C. D. Weis, T. Schenkel, and D. D. Awschalom, Physical Review Letters **101**, 117601 (2008).
[17]  E. Togan, Y. Chu, A. S. Trifonov, L. Jiang, J. Maze, L. Childress, M. V. G. Dutt, A. S. Sorensen, P. R. Hemmer, A. S. Zibrov, and M. D. Lukin, Nature **466**, 730 (2010).
[18]  T. M. Sweeney, C. Phelps, and H. L. Wang, Physical Review B **84**, 075321 (2011).
[19]  K. Bergmann, H. Theuer, and B. W. Shore, Reviews of Modern Physics **70**, 1003 (1998).
[20]  E. Togan, Y. Chu, A. Imamoglu, and M. D. Lukin, Nature **478**, 497 (2011).
[21]  C. Santori, P. Tamarat, P. Neumann, J. Wrachtrup, D. Fattal, R. G. Beausoleil, J. Rabeau, P. Olivero, A. D. Greentree, S. Prawer, F. Jelezko, and P. Hemmer, Physical Review Letters **97**, 247401 (2006).
[22]  P. R. Hemmer, A. V. Turukhin, M. S. Shahriar, and J. A. Musser, Optics Letters **26**, 361 (2001).

[23]   D. A. Golter, K. N. Dinyari, and H. L. Wang, Physical Review A **87**, 035801 (2013).
[24]   C. G. Yale, B. B. Buckley, D. J. Christle, G. Burkard, F. J. Heremans, L. C. Bassett, and D. D. Awschalom, Proceedings of the National Academy of Sciences of the United States of America **110**, 7595 (2013).
[25]   P. Tamarat, T. Gaebel, J. R. Rabeau, M. Khan, A. D. Greentree, H. Wilson, L. C. L. Hollenberg, S. Prawer, P. Hemmer, F. Jelezko, and J. Wrachtrup, Physical Review Letters **97**, 083002 (2006).
[26]   Y. Shen, T. M. Sweeney, and H. Wang, Physical Review B **77**, 033201 (2008).
[27]   K. M. C. Fu, C. Santori, P. E. Barclay, L. J. Rogers, N. B. Manson, and R. G. Beausoleil, Physical Review Letters **103**, 256404 (2009).
[28]   G. de Lange, Z. H. Wang, D. Riste, V. V. Dobrovitski, and R. Hanson, Science **330**, 60 (2010).
[29]   M. O. Scully and M. S. Zubairy, *Quantum optics* (Cambridge University Press, 1997).
[30]   See the supplement for a more detailed theoretical analysis of the experimental results.
[31]   P. Siyushev, H. Pinto, M. Voros, A. Gali, F. Jelezko, and J. Wrachtrup, Physical Review Letters **110**, 167402 (2013).

# Supplementary Information: Optically-driven nuclear-spin-selective Rabi oscillations of single electron spins in diamond

D. Andrew Golter and Hailin Wang

Department of Physics and Oregon Center for Optics
University of Oregon, Eugene, Oregon 97403, USA

**Discussion of the Theoretical Model**

To analyze the results of our optically driven Rabi oscillations and stimulated Raman adiabatic passage (STIRAP) experiments, especially the effects due to various broadening mechanisms, we performed a simulation using the model three level system shown in Fig. S1. We numerically solved the equation

$$\frac{\partial \rho}{\partial t} = -i[H, \rho] + D(\rho)$$

with

$$H = \begin{pmatrix} \Delta - \frac{\delta}{2} & 0 & \frac{\Omega_-}{2} \\ 0 & \Delta + \frac{\delta}{2} & \frac{\Omega_+}{2} \\ \frac{\Omega_-}{2} & \frac{\Omega_+}{2} & 0 \end{pmatrix}$$

and

$$D(\rho) = \begin{pmatrix} \Gamma \rho_{ee} & -\gamma_s \rho_{-+} & -\gamma \rho_{-e} \\ -\gamma_s \rho_{+-} & \Gamma \rho_{ee} & -\gamma \rho_{+e} \\ -\gamma \rho_{e-} & -\gamma \rho_{e+} & -2\Gamma \rho_{ee} \end{pmatrix}$$

where we have taken the rotating wave approximation and transformed into the rotating frame.

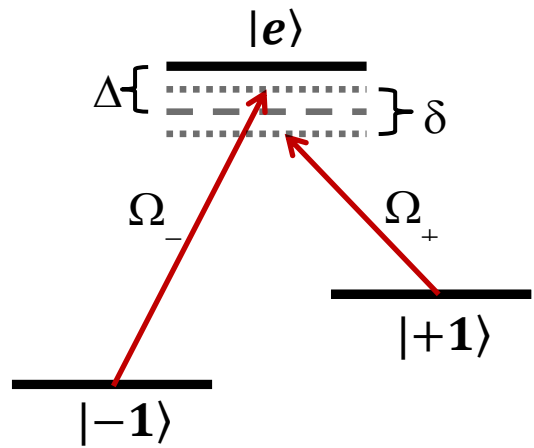

**FIG. S1** Energy level struture used in the model.

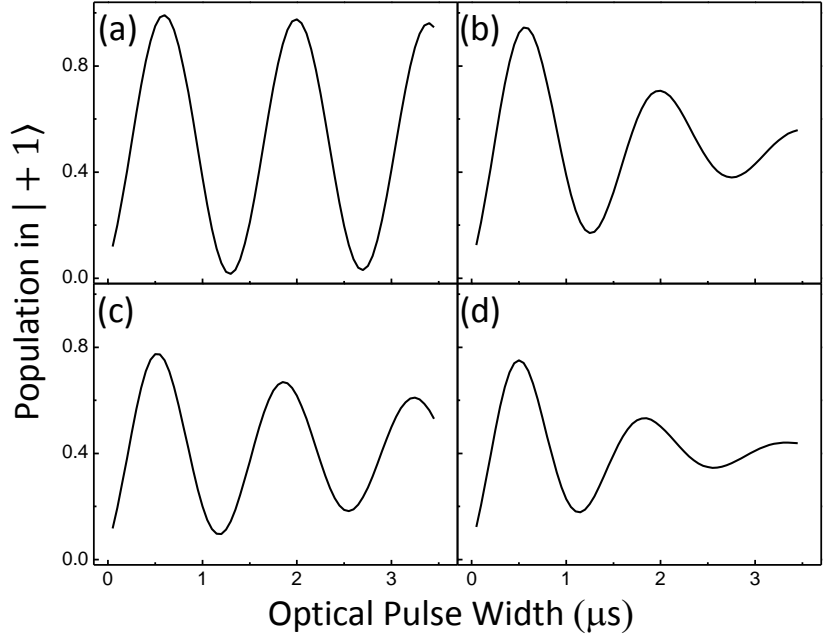

**FIG. S2. Simulation of the Rabi oscillation measurements.** (a) $\Delta$ held constant and $\delta = 0$. (b) Includes spectral diffusion. Sum over $\Delta$'s with $\delta = 0$. (c) Includes dephasing. Sum over $\delta$'s with $\Delta$ held constant. (d) Includes both spectral diffusion and dephasing. Sum over $\Delta$'s and $\delta$'s.

$\Delta$ is the average detuning, $\delta$ is the detuning from Raman resonance, $\Omega_+$ and $\Omega_-$ are the Rabi frequencies associated with each field, $\Gamma/2\pi = \gamma/2\pi = 7$ MHz, and $\gamma_s/2\pi = 1/(2\pi * T_2) = .8$ kHz. This model ignores decay out of the $\Lambda$-system, which is a reasonable approximation [1].

To model the Rabi oscillations shown in Fig. 1c of the main text, we set $\Delta/2\pi = 1.5$ GHz, $\delta/2\pi = 0$, and $\Omega_+/2\pi = \Omega_-/2\pi = 46$ MHz. The final population in level $|+1\rangle$ ($\rho_{++}$) is plotted as a function of pulse width (Fig. S2a). The small amount of decay is mostly due to optical pumping from direct excitation into $A_2$. In the experiment, the two hyperfine states for which the optical fields are not on Raman resonance will contribute to the background through pumping but not to the Rabi oscillation signal.

Each repetition of the experiment begins with off-resonant excitation which can cause spectral diffusion [2]. This acts as an inhomogeneous broadening of the $A_2$ transition. We model this by summing the results of our simulation over a range of $\Delta$'s, weighting them by a Gaussian with a FWHM/$2\pi$ = 500 MHz as estimated from PLE measurements. This produces the plot shown in Fig. S2b. Since the effective Rabi frequency depends on $\Delta$, the sum is over a range of Rabi frequencies resulting in an increase of the measured decay. If the spectral fluctuations of the NV optical transition were instead included as a part of the intrinsic decoherence rate, the calculation would predict effects of optical pumping that are much greater than those observed in the experiment.

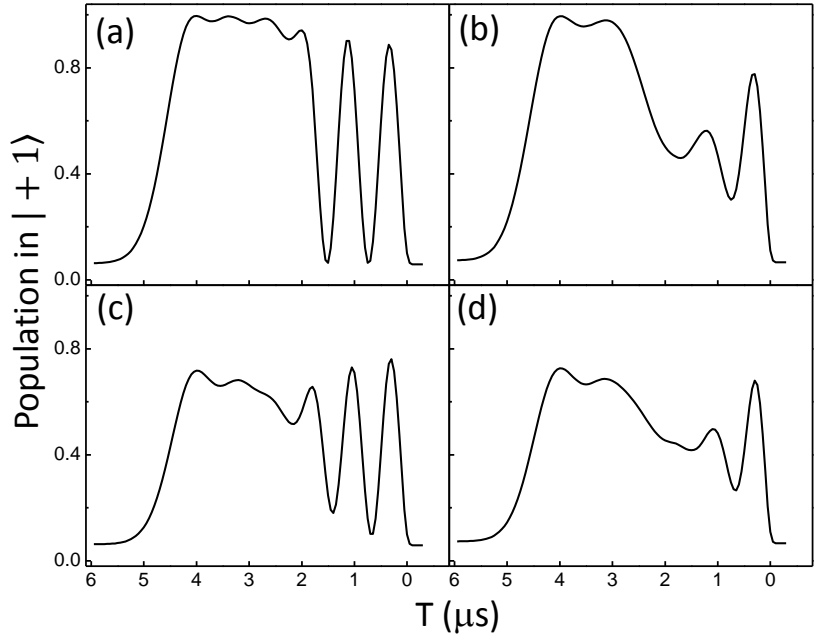

**FIG. S3. Simulation of the STIRAP/Rabi oscillation measurements.** (a) Δ held constant and $\delta = 0$. (b) Includes spectral diffusion. Sum over Δ's with $\delta = 0$. (c) Includes dephasing. Sum over $\delta$'s with Δ held constant. (d) Includes both spectral diffusion and dephasing. Sum over Δ's and $\delta$'s.

To describe the decay rate observed in our experiments, we also need to add the effects of dephasing between the $m_s$=+1 and $m_s$=-1 levels. From our measurement of Ramsey fringes we determined a dephasing time of $T_2^* = 1\mu s$. To model this, we sum over a range of $\delta$'s weighted by a Gaussian with a FWHM/$2\pi = 1$ MHz. Fig. S2c shows the results with Δ held constant and these dephasing effects included. Cases with a non-zero $\delta$ (both positive and negative) exhibit smaller oscillation amplitudes and increased Rabi frequencies. This means that the net effect of including dephasing is to produce oscillations with a smaller amplitude, a slightly shifted Rabi frequency, and an increased decay rate. Fig. S2d includes both the spectral diffusion and the dephasing.

The STIRAP/Rabi oscillation experiments were modeled in the same way, with the results shown in Fig. 4 of the main text. Now $\Delta/2\pi = .9$ GHz and $\Omega/2\pi = 48$ MHz were used. Given these parameters, optical pumping is expected to be a larger effect. Additionally, the process may no longer be completely nuclear spin dependent; however using a smaller $\Omega_R$ should recover this dependence. The scale of the simulated plots was adjusted to match that of the experiments so that the shapes could be easily compared.

As with Fig. S2, Figs. S3 shows the results of the simulation for (a) Δ constant with $\delta = 0$, (b) a sum over Δ's, (c) a sum over $\delta$'s, and (d) a sum over both Δ's and $\delta$'s. We see that including spectral diffusion has a relatively large effect on the Rabi oscillations, but a relatively small effect on the STIRAP. This is to be expected since the STIRAP process is relatively insensitive to Δ. The dephasing, on the other hand, has a smaller effect on the Rabi oscillations; and it reduces the overall efficiency of the STIRAP without significantly changing the adiabaticity.

**References**

[1]  E. Togan, Y. Chu, A. Imamoglu, and M. D. Lukin, Nature **478**, 497 (2011).

[2]  P. Tamarat, T. Gaebel, J. R. Rabeau, M. Khan, A. D. Greentree, H. Wilson, L. C. L. Hollenberg, S. Prawer, P. Hemmer, F. Jelezko, and J. Wrachtrup, Phys. Rev. Lett. **97**, 083002 (2006).